\documentclass[runningheads]{svmult}

\usepackage{makeidx}   
\usepackage{graphicx}  
\usepackage{subeqnar}  
\usepackage{multicol}  
\usepackage{physprbb}  
\makeindex             

\begin{document}

\title{AGB stars in the Local Group, and beyond}
\author{M.A.T. Groenewegen}
\institute{Instituut voor Sterrenkunde, KU Leuven, Celestijnenlaan 200B, \\ 
B-3001 Leuven, Belgium}

\maketitle

\vspace{-6mm}
\section{Introduction}
\label{intro}

Carbon stars are tracers of the intermediate age population in
galaxies. Either they are currently undergoing third dredge-up on the
(Thermal-Pulsing) Asymptotic Giant Branch (TP-AGB) -- the cool and
luminous N-type carbon stars (or the ``infrared carbon stars'', when
they are significantly reddened by circumstellar dust through
mass-loss) --, or have been enriched with carbon-rich material in a
binary system when the present-day white dwarf was on the AGB (the
carbon dwarfs and CH-stars. The R-stars may be the result of a
coalescing binary [McClure 1997]).

Similarly, there are ``intrinsic'' and ``extrinsic'' S-stars which
represent, respectively, the stars currently on the AGB, and the stars
polluted by a present-day white dwarf (e.g. Jorissen 2003).

Next to the classical oxygen-rich giants (of spectral type M and MS),
there are the OH/IR stars (which are obscured by circumstellar dust)
and the Barium-dwarfs and Barium-giants which are polluted by a
present-day WD (e.g. Jorissen 2003).

Since their spectral signature is very different from oxygen-rich and
S-type stars, it is relatively easy to identify carbon stars even at
large distances. In Sect.~\ref{NB} the main technique to identify
carbon stars is briefly discussed, and various recent surveys for
carbon stars in external galaxies are summarised. That section is an
update of reviews I gave at IAU Symposium 191 (Groenewegen 1999;
hereafter G99), and a Ringberg conference (Groenewegen 2002; hereafter
G02), and here I will only refer to new results published since
then. Please refer to G99, G02 and the similar review by Azzopardi
(1999) for the full story. Sections~\ref{NIR} and \ref{VAR} discuss
aspects related to NIR work and variability properties of AGB stars in
LG galaxies.


\section{Optical narrow-band imaging}
\label{NB}

This has been discussed in some detail in G99. Briefly, the most
effective method for large scale surveys of late-type M- and C-stars
uses typically two broad-band filters from the set $V,R,I$, and two
narrow-band filters near 7800 and 8100 \AA, which are centred on a
CN-band in carbon stars, and a TiO band in oxygen-rich stars,
respectively. In an [78-81] versus $[V-I]$ (or $[R-I]$) colour-colour
plot, carbon stars and late-type oxygen-rich stars clearly separate
redwards of $(V-I) \approx$ 1.6.  For an illustration of this, see
Cook \& Aaronson (1989) or Nowotny \& Kerschbaum (2002).

A caveat is that, unfortunately, not all groups adopt the same
``boxes'' in these diagrams to select M- and C-stars (e.g. compare
Nowotny et al. 2003 and Battinelli \& Demers et al. 2004b). 
Furthermore, in many cases, only the photometry is published for the
stars the respective authors consider to be the M- and C-stars, so it
is not possible to apply ones own selection criteria a-posteriori. 
Furthermore, one usually applies the same lower-limit on the broad
band colour ($(V-I)$, or $(R-I)$) to select M-stars (usually chosen to
correspond to M0 and later for solar metallicity) to C-stars, while it
is known (e.g.  Nowotny \& Kerschbaum 2002)) that the hottest C-stars
(spectral type C0) are bluer than this limit.  So, applying the same
lower-limit on the broad-band colour will bias against the hottest C-stars.

\begin{table}[t]
\caption[]{The carbon star census}
\setlength{\tabcolsep}{2.0mm}

\begin{flushleft}
\begin{tabular}{lrrrrll} \hline
Name      & D    & $M_{\rm V}$ & [Fe/H]     & $N_{\rm C}$$^{(b)}$ & Area$^{(c)}$ &  $N_{\rm M}$$^{(d)}$ \\ 
          & (kpc)& (mag)       &            &             & (kpc$^2$)   &   \\ \hline
M31       &  770 & --21.2      &   0.0      & 243         & 12.3 & 789 (5+) \\
Galaxy$^{(a)}$ &      & --20.9      &   0.0      &  81         & 1.00 & C/M $\approx$ 0.2  \\
M33       &  840 & --19.0      & --0.6      &  15         & 0.20 & 5 (5+), 60 (0+) \\
LMC       &   50 & --18.5      & --0.6      & 1045        & 4.8  & 1300 (5+) \\
          &      &             &            & 7750        & 220. &  \\
SMC       &   63 & --17.1      & --1.2      &  789        & 5.4  & 180 (5+)\\
          &      &             &            & 1707        & 12.2 &  \\
NGC 205   &  830 & --16.4      & --0.8      & 525         & 12.3 & 5830 (0+) \\
NGC 6822  &  490 & --16.0      & --1.2      & 904         & 4.5  & 941 (0+) \\
NGC 3109  & 1360 & --15.7      & --1.7      & 446         & 33   & 250 (0+) \\
NGC 185   &  620 & --15.6      & --0.8      & 145         & 4.5  & 850 (0+) \\
IC 1613   &  715 & --15.3      & --1.4      & 195         & 7.8  & 35 (5+), 300 (0+) \\
NGC 147   &  755 & --15.1      & --1.1      & 288         & 25.  & 1200 (0+) \\
SagDSph   &   28 & --15.0      & --0.5      & 26          & 7.2  &    \\
WLM       &  930 & --14.4      & --1.5      & 149         & 17.0 & 12 (0+) \\
Fornax    &  138 & --13.1      & --1.2      & 104         & 1.35 & 4 (5+), 25 (2+)\\
Pegasus   &  760 & --12.9      & --2.0      & 40          & 1.04 & 77 (0+) \\
SagDIG    & 1060 & --12.0      & --2.3      & 16          & 0.58 & 1 (0+) \\
Leo I     &  250 & --11.9      & --1.4      & 23          & 0.45 &  1 (5+), 15 (0+) \\
And I     &  790 & --11.8      & --1.4      & 0           & 0.33 & \\
And II    &  680 & --11.8      & --1.5      & 8           & 0.35 & 1 (0+) \\
And III   &  760 & --10.2      & --1.7      & 0           & 0.66 & \\
And V     &  810 &  --9.1      & --1.9      & 0           & 0.66 &  \\
And VI    &  775 & --11.3      & --1.7      & 1           & 0.41 &  \\
And VII   &  760 & --12.0      & --1.5      & 3           & 0.24 &  \\
DDO210    &  950 & --10.9      & --1.9      & 3           & 0.18 & 1 (0+) \\ 
Leo II    &  205 & --10.1      & --1.6      & 8           & 0.47 & \\
Cetus     &  775 & --10.1      & --1.7      & 1           & 1.1  &  \\
Sculptor  &   88 &  --9.8      & --1.5      & 8           & 0.65 & 40 (2+), 0 (5+)\\
Phoenix   &  405 &  --9.8      & --1.9      & 2           & 0.40 & \\
Tucana    &  870 &  --9.6      & --1.7      & 0           & 0.22 & \\
Sextans   &   86 &  --9.5      & --1.9      & (0)         &      & \\
Draco     &   79 &  --9.4      & --2.0      & 6           & 0.50 & \\ 
Carina    &   94 &  --9.4      & --1.8      & 11          & 0.31 & \\
Ursa Minor&   69 &  --8.9      & --1.9      &  7          & 0.58 & \\ 
NGC 2403  & 3390 & --20.4      &   0.0      &  4          & 2.0  & 7 (0+) \\
NGC 300   & 2170 & --18.7      & --0.4      & 16          & 3.2  & 23 (0+), 6 (5+) \\
NGC 55    & 1480 & --18.0      & --0.6      & 14          & 2.8  & 6 (5+) \\
\hline
\end{tabular}

$^a$. In the solar neighbourhood per kpc$^2$ (Groenewegen et al. 1992)

$^b$. Number of known carbon stars.

$^c$. Survey area at the assumed distance of the galaxy.

$^d$. Number of M-stars. 0+ indicates M0 and later, etc.

\end{flushleft}
\vspace{-7mm}
\label{TAB1}
\end{table}

\subsection{Surveys}
\label{surveys}

In this section the (recent, post-2002) surveys for carbon stars in external
galaxies are described.

\underline{M31}

\noindent
Battinelli et al. (2003) present results on a 2240 arcmin$^2$ area in
the SW disk of M31, identifying 945 carbon stars and estimating a C/M
ratio of 0.084. Previous work has been mentioned in G99 and G02.

\underline{NGC 205}

\noindent
Demers et al. (2003a) detect 289 carbon stars inside an 10 arcmin
ellipse around NGC 205, and estimate that the actual total number is
between 500 and 550, the difference being due to a lack of detections
near the centre (their figure 6) due to crowding and increased
photometric error. A C/M ratio of 0.09 is estimated.

\begin{figure}[t]
\includegraphics[width=\textwidth]{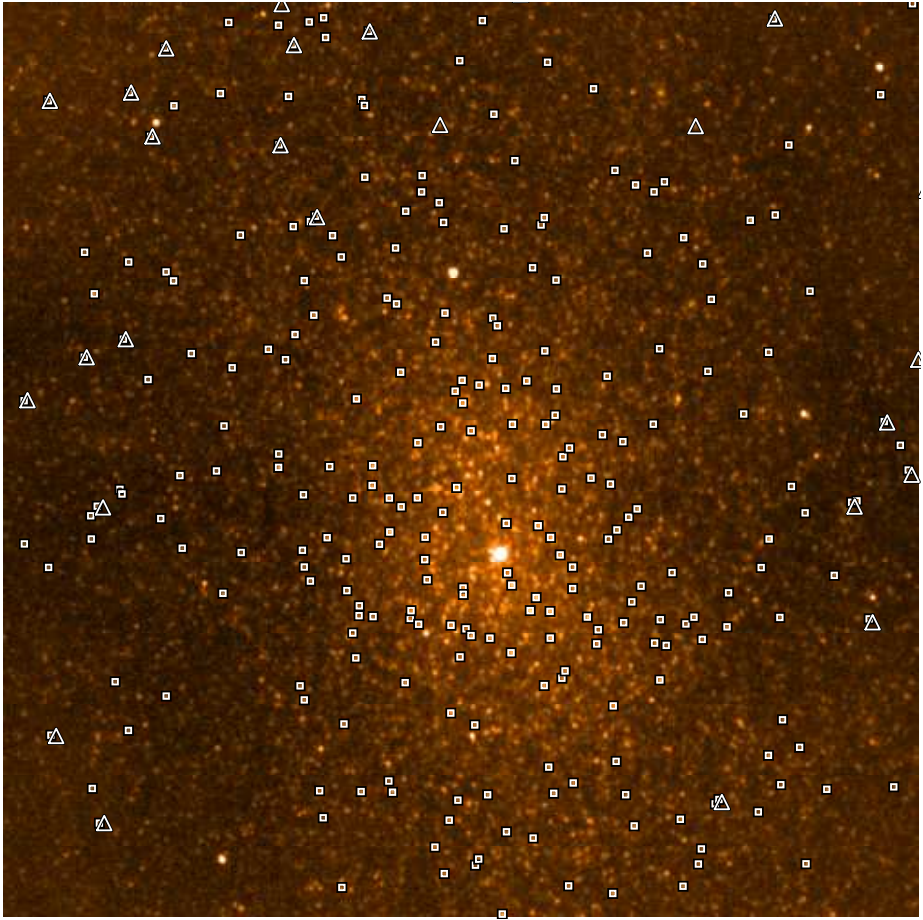}
\caption[]{NGC 205 in the $K$-band (Davidge 2003, FOV of 3.6
arcmin$^2$): known carbon stars from Demers et al. (triangles) and
objects with $(J-K)>1.5$ and $K < 17.2$ from Davidge (squares).}
\vspace{-4mm}
\label{N205}
\end{figure}

Davidge (2003) present $JHK$ results on a 3.6 arcmin$^2$ centred on
NGC 205, and he kindly provided the $K$-band fits image and object list. 
I derived the WCS information by identifying objects on a 2MASS
$K$-image resulting in astrometry with an rms of 0.21 and 0.30 arcsec
in RA and DEC using 18 objects over the entire field. Figure~\ref{N205} shows 
the $K$-image with plotted on top the known C-stars from Demers et al., 
and candidate AGB stars ($(J-K)>1.5$ and $K < 17.2$) from Davidge.

Even in the outer annulus, where Demers et al. suggest that their
survey is not hampered by crowding, there are many candidate AGB stars
based on $JK$ colours and magnitudes. Second, as both papers quote a
seeing of 0.7 arcsec, it illustrates the power of infrared observations.

Of the 25 known C-stars, 23 have a counterpart in the Davidge dataset
(although not all have $(J-K)>1.5$ and $K < 17.2$). The remaining two are
all visible on the $K$-image; one is on the edge of the frame, and the
second was missed in the source detection/extraction.

\begin{figure}[t]
\includegraphics[width=\textwidth]{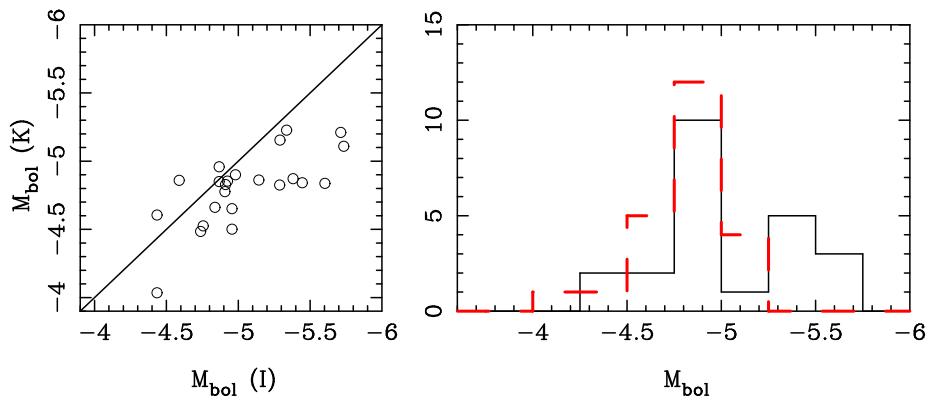}
\caption[]{Bolometric magnitudes for the 23 C-stars in common between
Demers et al. (solid histogram) and Davidge (dashed histogram) for NGC
205, based on $JK$ and $RI$ photometry and standard bolometric
correction formulae.  }
\vspace{-4mm}
\label{N205CC}
\end{figure}

One interesting exercise is to compare the bolometric magnitudes
derived from $JK$, with those derived from $RI$ photometry.  This
comparison is shown in Figure~\ref{N205CC} both on a star-by-star
basis and in the form of luminosity functions. Although these stars
are expected to be LPVs (hence variable) the data suggests possible
systematic effects, which warrants further investigation.


It might be recalled in this respect that one advantage of using $VI$
instead of $RI$ as broad-band colours is the fact the bolometric
correction to $I$ based on $(V-I)$ has much less scatter than the one
based on $(R-I)$ (Bessell \& Wood 1984).

\underline{NGC 3109}

\noindent
Demers et al. (2003b) present the first survey of this galaxy using the
narrow-band filter technique and detect 446 C-stars, and derive a C/M
ratio of about 1.8

\underline{NGC 147 \& NGC 185}

\noindent
Nowotny et al. (2003) and Battinelli \& Demers (2004b,c) independently
present data on these two companions to M31, while Harbeck et
al. (2004) observed NGC 147. Nowotny et al. find 146 C-stars and a C/M
ratio of 0.15 in NGC 147, and 154 and 0.089 in NGC 185, respectively,
over an un vignetted FOV of approximately 33 arcmin$^2$. Battinelli \&
Demers (2004c) find 288 C-stars and a C/M ratio of 0.24 for NGC 147
and Battinieli \& Demers (2004b) find 145 C-stars and a C/M ratio of
0.17 for NGC 185, both over an area of 1180 arcmin$^2$. Figure~3 in
Battinieli \& Demers (2004b) shows that most of their C-stars are
located in the main body of NGC 185, which is still well covered by
the smaller FOV of the Nowotny et al. observations. This is consistent
with the very similar numbers of C-stars discovered in both
surveys. On the contrary, the difference in number of C-stars detected
between the two datasets in the case of NGC 147 is--at least in
part--due to the difference in areal coverage of this galaxy and the
more extended distribution of the C-stars. Harbeck et al. (2004) find
155 C-stars in their 92 arcmin$^2$ FOV.

\underline{And {\sc iii, v, vi, vii}}

\noindent
Harbeck et al. (2004) also observed And {\sc iii, v, vi, vii} to
derive, respectively, 1, 0, 2, 5 C-stars, of which, respectively 0, 0,
1, 3 are believed to be genuine AGB C-stars, while the others are
suggested to be low-luminosity CH-stars.

\underline{WLM}

\noindent
Battinelli \& Demers (2004a) observe a 42 x 28 arcmin field centred
on WLM to find 149 C-stars, and to derive a C/M ratio of 12.4 $\pm 3.7$.

\underline{Cetus}

\noindent
Harbeck et al. (2004) also observed the Cetus dwarf spheroidal to find
1 genuine AGB C-star, and 2 CH-stars.


\begin{figure}[t]
\includegraphics[width=0.8\textwidth]{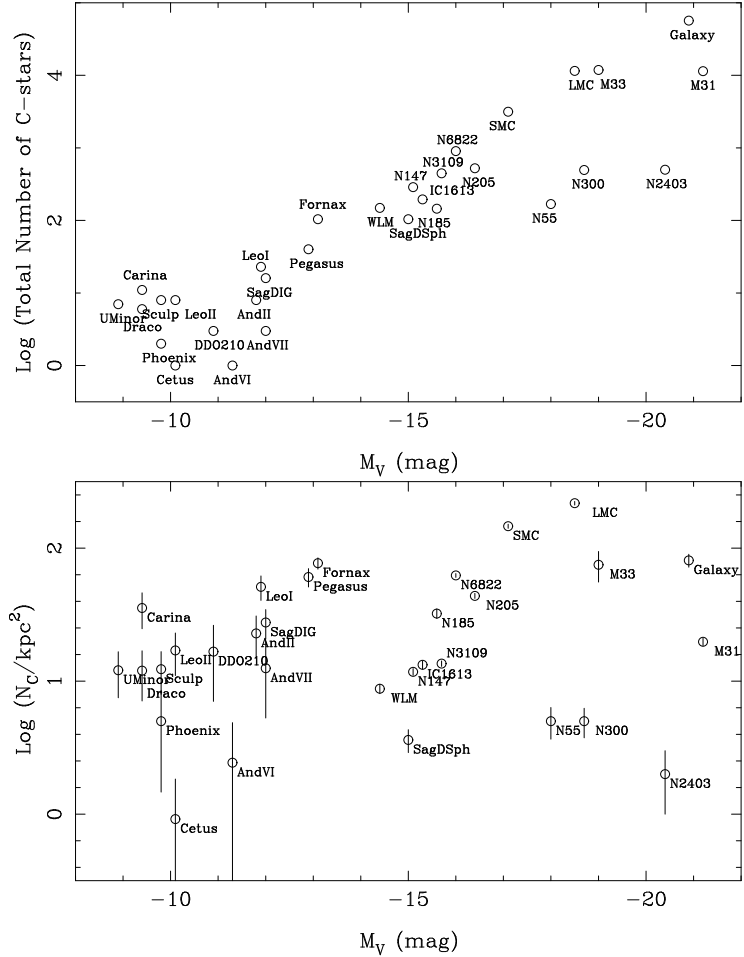}
\caption[]{Total number of carbon stars and surface density of carbon
stars versus $M_{\rm V}$. No correction for inclination effects has
been made.}
\vspace{-4mm}
\label{MON2}
\end{figure}

\begin{figure}[t]
\includegraphics[width=0.8\textwidth]{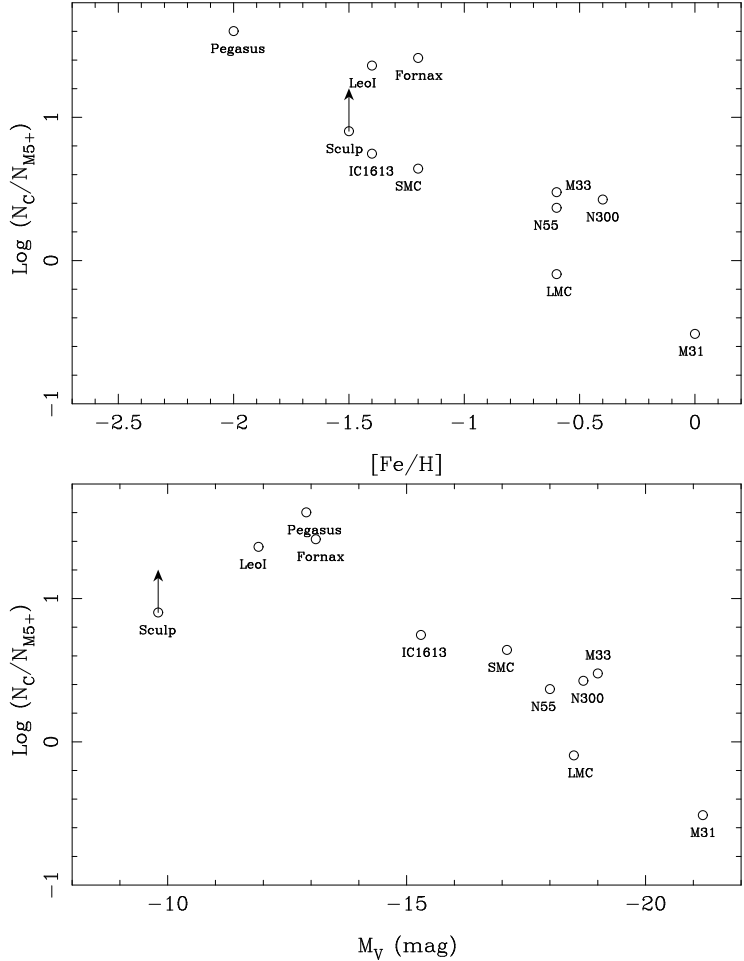}
\caption[]{Log (number carbon stars / number late M-stars) versus metallicity and $M_{\rm V}$. }
\vspace{-4mm}
\label{MON5}
\end{figure}

\begin{figure}[t]
\includegraphics[width=0.8\textwidth]{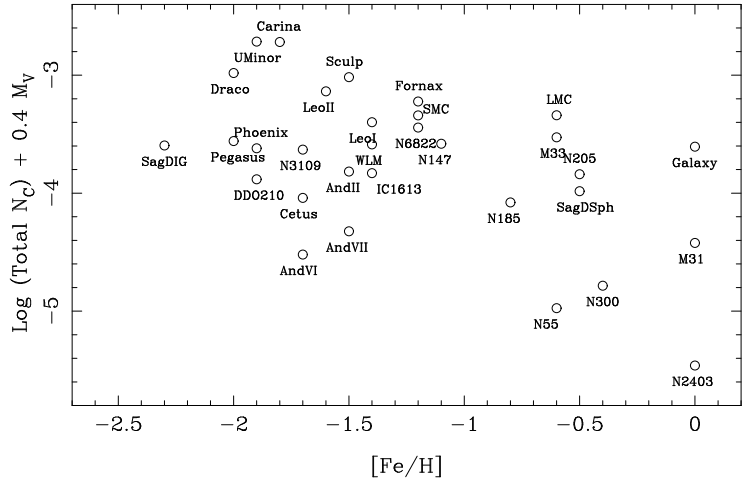}
\caption[]{Log (number of carbon stars / total visual luminosity) versus metallicity.}
\vspace{-4mm}
\label{MON6}
\end{figure}

\vspace{5mm}
\noindent
Table~\ref{TAB1} summarises the number of known carbon stars in
external galaxies. Local Group members not explicitly mentioned have
no published information on their C-star population. The last three
entries are galaxies outside the Local Group. Listed are the adopted
distance, absolute visual magnitude, metallicity (these three
parameters come from Grebel et al. (2003), Mateo (1998) and van den
Bergh (2000)), number of known carbon stars, the surface area on the
sky of the respective survey, and the number of late-type M-stars,
when known 


In the near future one may expect new results from the optical
narrow-band imaging technique for Carina, Sculptor, Phoenix
(Groenewegen et al.), IC 10 (Demers, Battinelli), and And {\sc ii},
M32, Leo {\sc i}, Leo {\sc ii}, Draco, Ursa Minor (Kerschbaum,
Nowotny, Olofsson, Schwarz). This last group also acquired funds to
put these filters on the 4.1m SOAR (Southern Astrophysical Research) telescope.

\begin{figure}[t]
\includegraphics[width=0.8\textwidth]{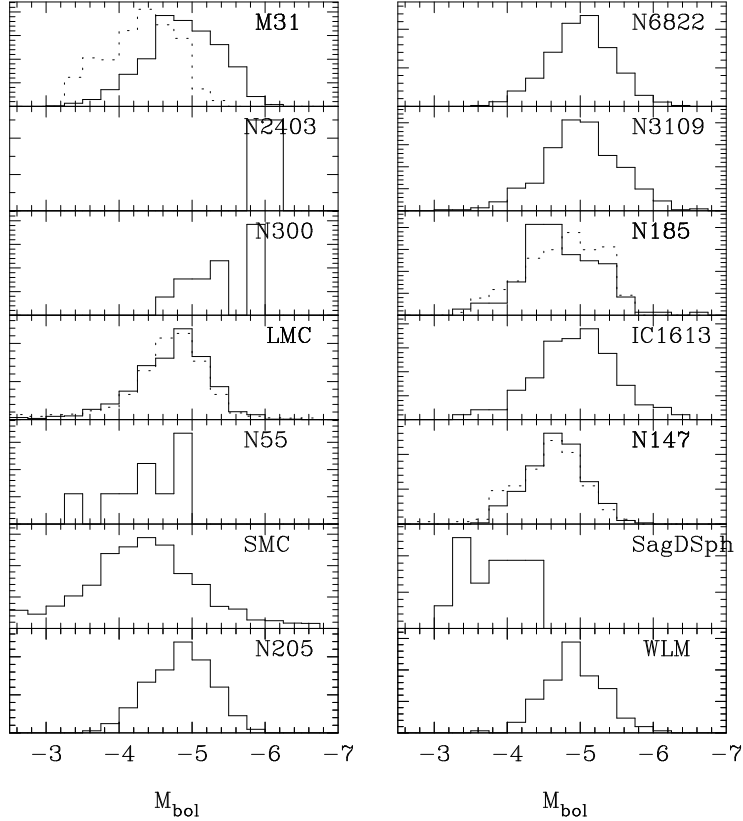}
\caption[]{Normalised carbon star luminosity functions, ordered by
decreasing $M_{\rm V}$ of the galaxies, ordered top to bottom, left to right.  
Number of C-stars used to calculate the LFs varies from galaxy to
galaxy (see Table~\ref{TAB1}.  In the case of SMC and LMC, the lowest
luminosity bin is cumulative.
For M31 data from Battinelli et al. (2003, solid) and Brewer et al.
(1995, dashed), for the LMC from Costa \& Frogel (1996, solid) and
Kontizas et al. (2001, dashed), for NGC 185 from Battinieli \& Demers
(2004b, solid) and Nowotny et al. (2003, dashed), for NGC 147 from
Battinelli \& Demers (2004c, solid) and Nowotny et al. (2003, dashed),
are plotted. For the LMC, NGC 147 and NGC 185 these independent LFs are
in good agreement. The difference for M31 appears to be real and
related to the fact that these LFs have been derived for (a) field(s) that
have different locations along the major axis of M31.
}
\vspace{-4mm}
\label{MON4a}
\end{figure}

\setcounter{figure}{6}
\begin{figure}[t]
\includegraphics[width=0.8\textwidth]{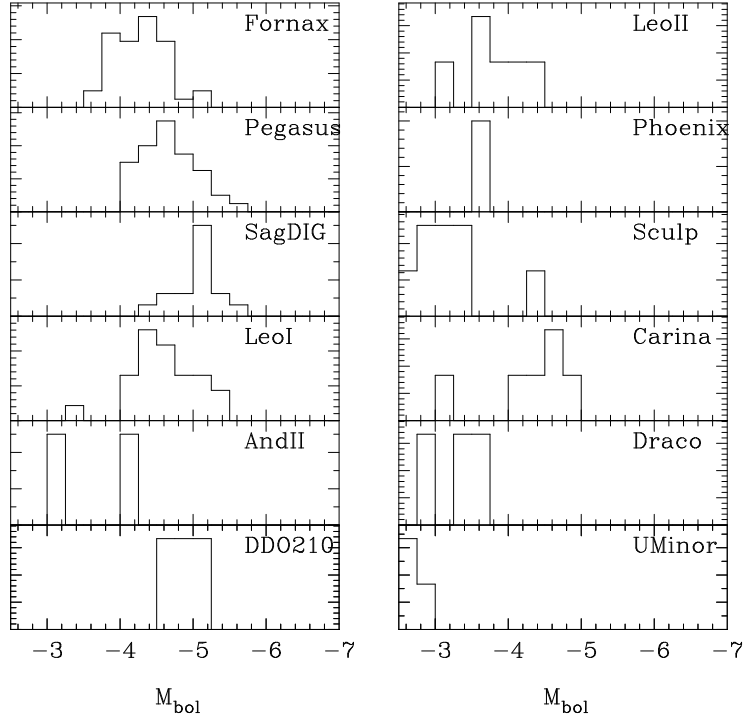}
\caption[]{Continued.}
\vspace{-4mm}
\label{MON4b}
\end{figure}


\section{Near-infrared results}
\label{NIR}

The comparison between the optical narrow-band imaging and
near-infrared observation of NGC 205 in the previous section already
demonstrated the power of the latter. A disadvantage of broad-band NIR
observations is that no distinction between M- and C-stars can be
made, although one often sees in the literature the assumption that
stars redder than $(J-K)_0 \sim 1.4-1.6$ are carbon stars. This is an
oversimplification, as dust-obscured (mass-losing) M-stars can have
red colours, and more importantly, hotter C-stars have NIR colours that
overlap with the photospheric colours of M-giants (e.g. figure~2 in G02).

The release of the 2MASS database (as well as DENIS data) has had a
large impact on the study of AGB stars in the Magellanic Clouds (MCs),
e.g. Cioni \& Habing (2003), but results have also been presented for
Fornax (Demers et al. 2002, G02), and the counterparts of optically
known C-stars at the time of the 2MASS {\it second data release}. More
recently, an undergraduate has investigated the 2MASS data of the {\it
all-sky release} for all LG galaxies within 1 Mpc (excluding MCs, M31,
M32), and retaining objects with $(J-K)_0 > 1.22$, appropriate $M_{\rm
K}$-range for AGB stars, and excluding known objects using the SIMBAD
database (Marescaux 2003). Table~\ref{TAB3} compares the number of
resulting candidate AGB stars (C-stars and dust-obscured M-stars) to
the number of known C-stars. For galaxies beyond $\sim$300 kpc the
number of candidates becomes very small and is limited by the survey
limit of 2MASS, restricting the candidates to the intrinsically most
luminous ones. For the more nearby galaxies the limiting magnitude of
2MASS is not a limiting factor and one observes that the number of
candidates is a non-negligible fraction of the number of known
C-stars, again indicating the power of IR observations to provide a
complementary picture to the optical narrow-band imaging.

In the near future deeper ground-based IR observations will become
available.  Several groups are observing or even monitoring LG
galaxies, e.g.  Fornax, Leo {\sc i}, Leo {\sc ii}, Carina, Sculptor,
Phoenix, Sextans by E.~Held using SOFI; Leo {\sc i}, Fornax, (all
smaller LG) using the SIRIUS camera on the IRSF (e.g. Menzies et
al. 2002), NGC 6822 using the new CPAPIR camera on the CTIO 1.5m (Demers,
Battinelli), Leo A, Leo {\sc i}, Leo {\sc ii}, Sex B, NGC 6822, Draco,
NGC 147, NGC 185 using INGRID on the WHT (Cioni \& Habing, e.g. this
conference).

Another exciting possibility is to use narrow-band filters in the
NIR. A first example was recently presented by \"Ostlin \& Mouhcine
(2004) who observed a 14 x 14 arcsec field in the metal-poor ([Fe/H]=
$-1.7$) galaxy IZw 18 (at a distance modulus of 30.5 !) using the
NICMOS F171M and F180M filters. In a carbon star the F171M filter is
located in the continuum while the F180M is centred on a deep CN band
at 1.77 $\mu$m. Combining this with a broad-band F160W there were able
to identify 5 C-stars, 1 M-type AGB and 20 supergiants.


\begin{table}
\caption{Candidate IR AGB stars based on 2MASS compared to the number of known C-stars}
\setlength{\tabcolsep}{2.3mm}
\begin{tabular}{lrrr lrrr} \hline
Name      & $D$  (kpc)   & $N_{\rm 2mass}$ &  $N_{\rm C}$ & 
Name      & $D$  (kpc)   & $N_{\rm 2mass}$ &  $N_{\rm C}$ \\  \hline  
Ursa Minor&   69 &  8  &   7  & NGC 6822  &  490 &  6  & 904  \\
Sculptor  &   79 &  8  &   8  & LGS 3     &  620 &  1  &   ?  \\
Draco     &   82 &  6  &   6  & IC 10     &  660 &  9  &   ?   \\
Sextans   &   86 & 10  &  (0) & IC 1613   &  720 &  3  & 195   \\
Carina    &  100 &  6  &  11  & NGC 147   &  755 &  1  & 288  \\
Fornax    &  138 & 34  & 104  & And I     &  790 &  1  &   0 \\
Leo II    &  205 &  1  &   8  & Leo A     &  800 &  2  &   ?   \\
Leo I     &  250 &  2  &  23  & Tucana    &  870 &  1  &   0   \\ 
\hline
\end{tabular}
\label{TAB3}
\end{table}

\section{Variability}
\label{VAR}

One of the characteristics of AGB stars is that they are variable on
different timescales and amplitudes, as has been clearly revealed by
the mircolensing surveys of the Magellanic Clouds and Galactic Bulge
(e.g. Wood et al. 1999, Wood 2000). It has been found that variable
AGB stars occupy different sequences (usually labelled ABCD) in
period-luminosity diagrams, with the large amplitude Mira variables on
sequence C.

In G02, some earlier work on (candidate) variable AGB stars in the
SagDSph, Fornax and IC 1613 was reported. Since then, Snigula et
al. (2004) mention 16 candidate LPVs (Long Period Variables) in Leo A,
and 5 in GR 8, Gallart et al. (2004) propose 6 LPV candidates in
Phoenix, and Rejkuba (2004) present a $K$-band $PL$-relation for 240
well defined Miras in NGC 5128.

\section{Discussion}

Figures~\ref{MON2}, \ref{MON5}, \ref{MON6}, \ref{MON4a}, \ref{MON4b}
are updates of those in G02, and for lack of space I refer to the last
section in G02 for additional details.

Figure~\ref{MON2} shows the number of carbon stars in a galaxy versus
$M_{\rm V}$ represented in two ways. First, the total number of
C-stars in a galaxy was estimated by simply multiplying the known
number by the ratio of total surface area of a galaxy to the survey
area. As for some galaxies the survey area is less than a few percent,
this correction factor can be quite large (and uncertain).  To
circumvent this, the bottom panel shows the surface density of carbon
stars in the particular survey. The drawback of this approach is that
it does not take into account the spatial variation of the density of
carbon stars within a galaxy. In neither approach did I correct for
the fact that we do not see these galaxies face-on. Some interesting
things can be noticed. There is a clear relation between the
(estimated) total number of C-stars and $M_{\rm V}$, and there seems
to be a maximum surface density of about 200 kpc$^{-2}$ averaged over
a galaxy. In both plots NGC 55, 300 and 2403 are clear outliers. These
are the most distant galaxies surveyed, and one might suppose that the
surveys have been incomplete. For NGC 55 the explanation probably lies
as well in the fact that we see this galaxy almost disk-on, and so
both the total number as well as the surface density have been
underestimated. Reddening within the galactic disk of the galaxy can
also play a role. For NGC 2403 the small number of carbon stars lies
in the fact that the survey has been incomplete. All 4 known C-stars
have luminosities that are much higher than the average in galaxies
for which we know the LF in more detail. The same is true for NGC 300.

Figure~\ref{MON5} shows the ratio of C- to late M-stars. The
interpretation of the well known trend is that a star with a lower
metallicity needs fewer thermal pulses to turn from an oxygen-rich
star into a carbon star.

Figure~\ref{MON5} shows the ratio of the total estimated number of
carbon stars over the visual luminosity of the galaxy. Most of the
galaxies scatter between a value of --3 and --4, with a few outliers
which are the same as noticed in Fig.~\ref{MON2}.

Figures~\ref{MON4a} and \ref{MON4b} shows the C-stars bolometric LF
for the galaxies for which it could be constructed. The data show that
in well populated LFs, the mean $M_{\rm bol}$ is between --4 and
--5. It also shows that the mean in NGC 300 and NGC 2403 is much
higher. Unless one would invoke a large uncertainty in the distance or
a burst of recent star formation, the most natural explanation lies in
the incompleteness of the surveys in these distant galaxies.  Finally,
the data shows that in the fainter galaxies the mean magnitude
increases and that a fair number of C-stars are of the low-luminosity type. 

\noindent
In a recent paper Mouhcine \& Lan\c{c}on (2003) present evolutionary
population synthesis models, including chemical evolution, with
special focus on intermediate age populations. Their models are the
first that are able to account for the observed trend in
Fig.~\ref{MON5} adopting `typical' SFH for Sa, Sb, Sc and Irr Hubble
type galaxies.  The AGB phase is included through a semi-analytical
treatment of the third dredge-up, with efficiency parameters set to
values that have been determined in other studies to fit the LMC
carbon star LF and C/M ratio.

\vspace{-5mm}
\section{Conclusion}

In principle, the overall carbon star LF and C/M ratio contains
information about the star-formation rate history from, say, 10 Gyr
ago (the low-luminosity C-stars in binaries) to a few-hundred Myr ago
(the high luminosity tail of the LF). Its a challenge to theoretical
models to use these constraints together with other data to derive the
chemical evolution and star formation history of these galaxies. The
models of Mouhcine \& Lan\c{c}on represent a first successful step in
this direction. \\

\vspace{-5mm}
\section*{Acknowledgments}

I would like to thank Tim Davidge for providing the $K$-band fits
image and his list of detections for NGC 205, and Pierre Royer (KUL) for
helping in deriving the WCS parameters.




\end{document}